\documentclass[aps,prl,prabib,twocolumn,
showpacs,preprintnumbers,amsmath,amssymb,floatfix,aps]{revtex4}
\usepackage{graphicx}% Include figure files \usepackage{dcolumn}
\begin{document}

\title{Superfluid transition in a Bose gas with correlated disorder}
\author{S. Pilati and S. Giorgini}
\affiliation{Dipartimento di Fisica, Universit\`a di Trento
and CNR-INFM BEC Center, I-38050 Povo, Trento, Italy}
\author{N. Prokof'ev}
\affiliation{Department of Physics, University of Massachusetts, Amherst, MA 01003, USA}
\affiliation{Russian Research Center ``Kurchatov Institute'', 123182 Moscow, Russia}

\begin{abstract}
The superfluid transition of a three-dimensional gas of hard-sphere bosons in a disordered medium is studied using quantum Monte Carlo methods. Simulations are performed in continuous space both in the canonical and in the grand-canonical ensemble. At fixed density we calculate the shift of the transition temperature as a function of the disorder strength, while at fixed temperature we determine both the critical chemical potential and the critical density separating normal and superfluid phases. In the regime of strong disorder the normal phase extends up to large values of the degeneracy parameter and the critical chemical potential exhibits a linear dependence in the intensity of the random potential. The role of interactions and disorder correlations is also discussed.
\end{abstract}

\maketitle

The interplay between superfluidity, interactions and disorder in quantum degenerate Bose systems (the so-called dirty boson problem) is a central topic in condensed matter physics, many aspects of which are still unsolved and under scrutiny.
Since the seminal work by Fisher {\it et al.}~\cite{FWGF}, the general understanding is that interactions are essential to stabilize the system and that superfluidity is lost for strong enough disorder, leading to a normal phase which at low temperatures is identified with the Bose glass phase. However, a quantitative description in terms of the relevant parameters of the random potential and other matters, such as the critical behavior and the role of dimensionality, are still open issues.

On the experimental side a large body of work was devoted to $^4$He adsorbed in porous media, such as Vycor glass and aerogels~\cite{Reppy,Glyde}. These studies investigated the behavior of the heat capacity and of the superfluid response~\cite{Reppy}, as well as the dynamic structure factor~\cite{Glyde} as a function of temperature and filling. However, no clear evidence was observed of a compressible Bose glass phase. More recently the dirty boson problem has been addressed using ultracold atoms, which offer unprecedented control and tunability of the disorder parameters and of the interaction strength. Interaction effects were studied in disordered optical potentials~\cite{EXP1}, even though the main effort has been given so far to the suppression of diffusion for non-interacting particles (Anderson localization)~\cite{EXP2}.

Many relevant theoretical contributions are based on quantum Monte Carlo simulations of the Bose-Hubbard Hamiltonian with disorder\cite{MCBoseHubbard}. In this lattice model the physical scenario is more involved than in continuous space because of the role played by commensurability and of the existence of the interaction driven phase transition to the Mott insulating state. Other theoretical approaches make use of mean-field approximations~\cite{MField,Lopatin} and are not reliable in the regime of strong disorder.

\begin{figure}
\begin{center}
\includegraphics[width=8cm]{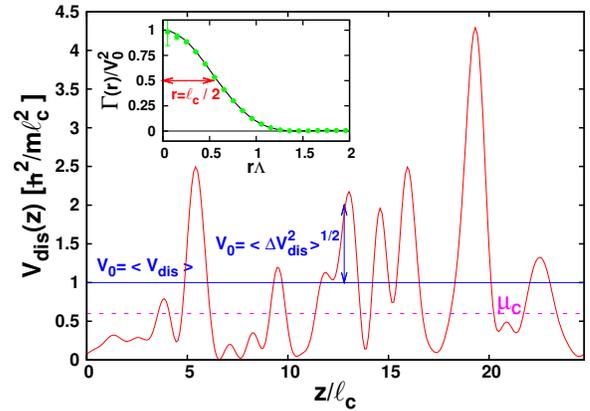}
\caption{(color online). Typical shape of the speckle potential $V_{\text{dis}}$, with averaged value $V_0=\hbar^2/m\ell_c^2$, shown in the direction (0,0,1) of the simulation box. We also show schematically the value of the critical chemical potential $\mu_c$. Inset: radial dependence (in units of the inverse momentum cut-off $\Lambda$) of the disorder spatial autocorrelation function $\Gamma$. The solid (black) line refers to an average over many realizations of the random field, the (green) symbols correspond to a single realization.}
\label{fig1}
\end{center}
\end{figure}

In this Letter we report on a path-integral Monte Carlo (PIMC) study of an interacting Bose gas in the presence of correlated disorder produced by 3D optical speckles. This random potential is relevant for experiments and allows for an independent tuning of intensity and correlation length. By increasing the disorder strength, we find a sizable reduction of the superfluid transition temperature and the shift is larger for weaker interactions. We map out the normal to superfluid phase diagram, both in the chemical potential vs. disorder and in the density vs. disorder plane. For strong disorder and in the presence of small but finite interactions, the critical chemical potential varies linearly with the disorder intensity and is essentially independent of temperature and interaction strength, in agreement with the existence of a mobility edge separating localized from extended states. In this regime and for chemical potentials below the critical value, the equilibrium state is a highly degenerate normal gas which is expected to correspond to the Bose glass phase.
We consider a system of $N$ identical particles of mass $m$ subject to the random field $V_{\text{dis}}$ and interacting with a short-range pairwise potential. The Hamiltonian is given by:
\begin{equation}
H=\sum_{i=1}^N \left(-\frac{\hbar^2}{2m}\nabla_i^2+V_{\text{dis}}({\bf r}_i)\right)+\sum_{i<j}V(|{\bf r}_i-{\bf r}_j|) \;.
\label{hamiltonian}
\end{equation}
The interatomic potential is modeled by a hard-sphere interaction: $V(r)=+\infty$ if $r<a$ and zero otherwise, where the hard-sphere diameter $a$ corresponds to the $s$-wave scattering length. The system is in a cubic box of volume $V=L^3$ with periodic boundary conditions. Disorder is modeled by an isotropic 3D speckle potential defined as follows~\cite{Huntley}:
\begin{equation}
V_{\text{dis}}({\bf r})=V_0\biggl|\frac{1}{V} \int d{\bf k} \tilde{\varphi}({\bf k})W({\bf k}) e^{i{\bf k}\cdot{\bf r}}\biggr|^2 \;,
\label{Vdis}
\end{equation}
where $V_0$ is a positive constant and $\tilde{\varphi}({\bf k})=\int d{\bf r} \varphi({\bf r}) e^{-i{\bf k}\cdot{\bf r}}$ is the Fourier transform of the complex field $\varphi({\bf r})$, whose real and imaginary part are independent random variables sampled from a gaussian distribution with zero mean and unit variance. The function $W({\bf k})$ is a low-wavevector filter defined as: $W({\bf k})=1$ if $k<\pi\Lambda$ and zero otherwise. The random potential in Eq.~\eqref{Vdis} is positive definite and the probability distribution of its intensities is given by the normalized exponential law $P(V_{\text{dis}})=e^{-V_{\text{dis}}/V_0}/V_0$. If the volume $V$ is large enough the disorder $V_{\text{dis}}$ is expected to be self-averaging, {\it i.e.} spatial averages coincide with averages over different realizations, and one has $V_0=\langle V_{\text{dis}}\rangle=1/V\int d{\bf r}V_{\text{dis}}({\bf r})$. The mean square displacement is also determined by the same energy scale: $V_0=\sqrt{\langle V_{\text{dis}}^2\rangle-\langle V_{\text{dis}}\rangle^2}$. The correlation length $\ell_c$ is defined from the spatial autocorrelation function, $\Gamma(r^\prime)=\langle V_{\text{dis}}({\bf r})V_{\text{dis}}({\bf r}+{\bf r}^\prime)\rangle-\langle V_{\text{dis}}\rangle^2$, as the length scale for which $\Gamma(\ell_c/2)=\Gamma(0)/2$. We find the following relation between the correlation length and the wave-vector cutoff $\Lambda$: $\ell_c=1.1/\Lambda$.
The length scale $\ell_c$ is typically $\sim 100$ times larger than the hard-sphere diameter $a$, allowing for a wide range of disorder intensities where interaction effects are well described by the $s$-wave scattering length and the details of the interatomic potential are irrelevant. The typical box size used in the simulations ranges from $L\sim20\ell_c$ to $L\sim50\ell_c$. An indication of self-averaging of disorder for these values of $L$ is provided by the inset of Fig.~\ref{fig1}, where we show the comparison between the autocorrelation function $\Gamma$ averaged over many realizations of the random potential and the one corresponding to a single realization. The typical shape of the speckle potential $V_{\text{dis}}$ is also shown in Fig.~\ref{fig1}: typical wells have size $\ell_c$ and depth $V_0$. We notice that standard experimental realizations of optical speckles are 2D, {\it i.e.} the speckle pattern lies in the plane perpendicular to the propagation of the laser beam. We consider instead a 3D pattern, having the same correlation length in the three spatial directions.

The energy $\hbar^2/m\ell_c^2$, associated with the correlation length $\ell_c$, and $V_0$ provide the two relevant energy scales for the disorder potential. In particular, if $V_0\gg\hbar^2/m\ell_c^2$ the random potential is classical in nature, with typical wells that are deep enough to sustain many single-particle bound states. The opposite regime, $V_0\ll\hbar^2/m\ell_c^2$, corresponds instead to quantum disorder, where typical wells of size $\ell_c$ do not have bound states and these can be supported only by rare wells of size much larger than $\ell_c$ or with depth much larger than $V_0$.

The outcomes of PIMC simulations consist of unbiased estimates of thermal averages of physical quantities, using the  many-particle configurations ${\bf R}=({\bf r}_1,...,{\bf r}_N)$ sampled from a probability distribution proportional to the density matrix $\rho({\bf R},{\bf R},T)=\langle{\bf R}|e^{-H/k_BT}|{\bf R}\rangle$ at the temperature $T$. In the present study we are interested in the superfluid density $\rho_s$, obtained from the winding number estimator~\cite{Ceperley}, and in the one-body density matrix (OBDM) $n_1(r)$, whose long-range behavior defines the condensate density $n_0=\lim_{r\to\infty}n_1(r)$. Our simulations are based on the worm algorithm~\cite{Boninsegni}, which allows for an efficient sampling of permutation cycles, and on the pair-product decomposition which is well suited for studies of dilute systems~\cite{PGP}. We perform calculations both in the canonical (at fixed density $n$) and in the grand-canonical ensemble (at fixed chemical potential $\mu$)~\cite{Boninsegni}.

\begin{figure}
\begin{center}
\includegraphics[width=8cm]{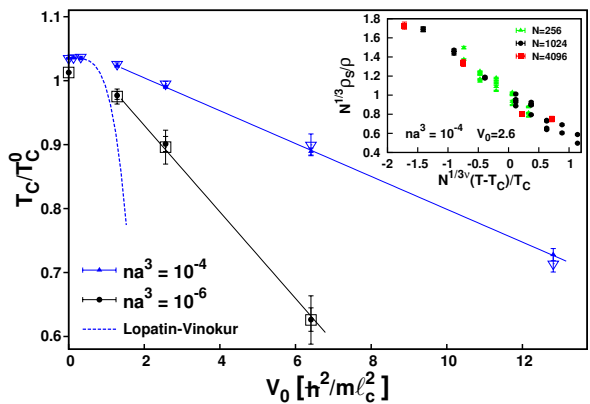}
\caption{(color online). Superfluid transition temperature as a function of the disorder strength for two values of the gas parameter $na^3$. Open and solid symbols refer respectively to $T_c$ determined from the superfluid and from the condensate fraction. The dashed line is the prediction of Ref.~\cite{Lopatin} at $na^3=10^{-4}$ shifted by $(T_c-T_c^0)$ in the absence of disorder. Inset: scaling behavior of the superfluid density for different system sizes and different realizations of disorder.}
\label{fig2}
\end{center}
\end{figure}

\begin{figure}
\begin{center}
\includegraphics[width=8cm]{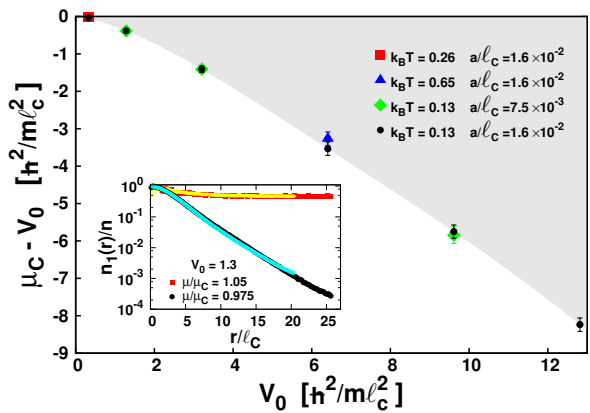}
\caption{(color online). Critical chemical potential (shifted by $V_0$) as a function of the disorder strength for different values of temperature (in units of $\hbar^2/m\ell_c^2$) and scattering length. The grey shaded area denotes the superfluid phase. Inset: spatial dependence of the OBDM for two values of the chemical potential slightly below and above $\mu_c$. Here $k_BT=0.13\hbar^2/m\ell_c^2$ and $a/\ell_c=0.016$. Two different system sizes are used to check the role of finite-size effects.}
\label{fig3}
\end{center}
\end{figure}

\begin{figure}
\begin{center}
\includegraphics[width=8cm]{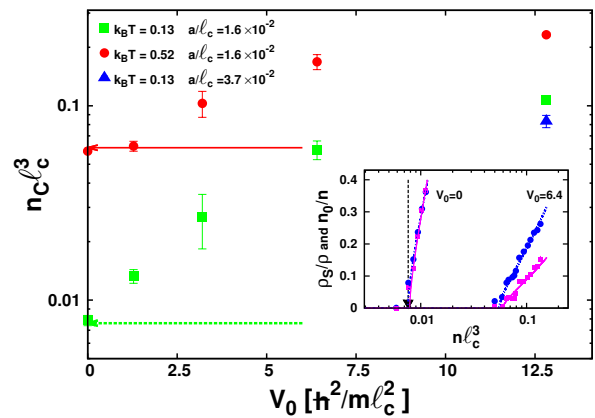}
\caption{(color online). Critical density as a function of the disorder strength for different values of temperature (in units of $\hbar^2/m\ell_c^2$) and scattering length. The horizontal arrows indicate the critical value $n_c^0$ of the non-interacting gas. Inset: Density dependence of $\rho_s/\rho$ [(pink) squares] and $n_0/n$ [(blue) circles] for the values $V_0=0$ and $V_0=6.4\hbar^2/m\ell_c^2$ of the disorder strength. Here $k_BT=0.13\hbar^2/m\ell_c^2$ and $a/\ell_c=0.016$. The vertical arrow indicates the corresponding value of the degenerate density $n_c^0$.}
\label{fig4}
\end{center}
\end{figure}

We are now in a position to discuss our results. First we discuss the simulations carried out at fixed density. The scattering length and the disorder correlation length are also kept fixed and for the latter we choose the value $n\ell_c^3=0.24$, such that there is typically one particle in each small sphere of radius $\ell_c$: $n4\pi\ell_c^3/3\simeq1$. Results for the transition temperature as a function of disorder strength are shown in Fig.~\ref{fig2} for two values of the gas parameter $na^3$. The transition temperature $T_c$ is expressed in units of $T_c^0=(2\pi\hbar^2/mk_B)[n/\zeta(3/2)]^{2/3}$, the critical temperature of the non-interacting gas with $\zeta(3/2)\simeq2.612$, and the results in the absence of disorder are taken from Ref.~\cite{PGP}. At $na^3=10^{-4}$, there is no appreciable change for $V_0\lesssim 1$ compared to $T_c$ in clean systems. For larger intensities we find a sizable shift that is well described by a linear dependence in $V_0$. For a given strength $V_0$ the reduction of the transition temperature is enhanced for smaller values of the gas parameter, consistently with the instability of the ideal Bose gas in the presence of disorder~\cite{GPS}. The value of $T_c$ is extracted from the results of the superfluid fraction $\rho_s/\rho$ ($\rho=mn$ is the total mass density), corresponding to systems with different particle number $N$, using the scaling ansatz
\begin{equation}
N^{1/3}\rho_s(t,N)/\rho=f(tN^{1/3\nu})= f(0)+f^\prime(0)tN^{1/3\nu}+... \;.
\label{scaling}
\end{equation}
Here, $t=(T-T_c)/T_c$ is the reduced temperature, $\nu$ is the critical exponent of the correlation length $\xi(t)\sim t^{-\nu}$, and $f(x)$ is a universal analytic function, which allows for a linear expansion around $x=0$. The validity of the scaling behavior (\ref{scaling}) is presented in the inset of Fig.~\ref{fig2}, where the effect of different realizations of the random potential is also shown. The quantity $N^{(1+\eta)/3}n_0/n$, involving the condensate fraction $n_0/n$ and the correlation function critical exponent $\eta=0.038$ of the XY-model universality class, is also expected to obey a scaling relation of the form (\ref{scaling}). For all reported disorder strengths $V_0$, the extracted value of the critical exponent $\nu$ is compatible with the result $\nu=0.67$ corresponding to clean systems~\cite{PGP}. It is worth noting that the values of $T_c$, obtained from the scaling law of the superfluid $\rho_s/\rho$ and of the condensate fraction $n_0/n$, coincide within our statistical uncertainty (see Fig.~\ref{fig2}). In Ref.~\cite{Lopatin} the shift $\delta T_c=T_c-T_c^0$ of the superfluid transition temperature is calculated using a perturbative approach for the $\delta$-correlated disorder $\langle\Delta V_{\text{dis}}({\bf r})\Delta V_{\text{dis}}({\bf r}^\prime)\rangle=\kappa\delta({\bf r}-{\bf r}^\prime)$, where $\Delta V_{\text{dis}}({\bf r})=V_{\text{dis}}({\bf r})-\langle V_{\text{dis}}\rangle$. The $T_c$ shift is
found to be quadratic in $\kappa$, implying for our speckle potential that $\delta T_c/T_c^0=(m^2V_0^2\ell_c^3/\sqrt{na}\hbar^4)^2/[2(12\log2)^3]$, where we used a gaussian fit to the radial dependence of the autocorrelation function $\Gamma$ and considered the limit $\ell_c\to0$. We report this prediction in Fig.~\ref{fig2} (we also add the interaction contribution not accounted for by Ref.~\cite{Lopatin}, so that in the clean case an exact result is reproduced). Our data
in the regime of very weak disorder do not have enough precision to allow for a quantitative comparison
and diverge from the theory before $\delta T_c/T_c^0$ becomes appreciable.
The effect of disorder on the critical temperature of a hard-sphere gas was also investigated using PIMC methods in Ref.~\cite{Gordillo} where, however, no significant reduction of $T_c$ was reported. For stronger intensities of disorder, the calculation of $T_c$ becomes increasingly difficult, since the dependence on the realization gets more important and larger systems are needed in order to have a satisfactory self-averaging of the random potential.

In Fig.~\ref{fig3} we report results for the critical chemical potential $\mu_c$ obtained from calculations carried out in the grand-canonical ensemble. A small change of $\mu$ around $\mu_c$ translates into a drastic change in the long-range behavior of the OBDM (see inset of Fig.~\ref{fig3}): for $\mu<\mu_c$ the OBDM decays to zero and corresponds to a normal phase, for $\mu>\mu_c$ the OBDM reaches a constant value characteristic of the superfluid state. If interactions are small but finite, we also find that the value of $\mu_c$ is essentially insensitive to a change of temperature and of interaction strength. For weak disorder, this result is accompanied by a very small critical density (see Fig.~\ref{fig4}) and corresponds to a renormalization of $\mu_c$ due to disorder in an extremely dilute gas. For strong disorder, it is instead consistent with the picture of a mobility edge, which depends only on the parameters of the random potential and separates localized single-particle states from extended ones. In this latter regime we find a linear dependence of $\mu_c$ as a function of $V_0$, in agreement with the qualitative $T=0$ prediction of Refs.~\cite{Shklovskii,Nattermann} in the case of classical disorder.

Finally we analyze the dependence of the critical density $n_c$ on the intensity of the random potential. The calculations are carried out in the canonical ensemble at fixed temperature and scattering length. The method used to determine $n_c$ is shown in the inset of Fig.~\ref{fig4}. For a given value of $V_0$ one increases the density and calculates the superfluid $\rho_s/\rho$ and the condensate fraction $n_0/n$. The results are then fitted by a power-law dependence $\rho_s/\rho\sim(n-n_c)^\nu$ and $n_0/n\sim(n-n_c)^{\nu(1+\eta)}$ for $n>n_c$, where the proportionality coefficients are expected to be non-universal parameters. In the inset of Fig.~\ref{fig4} we show the results corresponding to a configuration without disorder ($V_0=0$) and with strong disorder ($V_0=6.4\hbar^2/m\ell_c^2$). The reported values are averaged over a few realizations of the random potential and their scatter gives an idea of the relevance of this effect. For the small value of the scattering length used here, the critical density at $V_0=0$ coincides with the non-interacting result $n_c^0=\zeta(3/2)(mk_BT/2\pi\hbar^2)^{3/2}$, while for the large $V_0$ one finds that $n_c$ is about a factor of eight greater than $n_c^0$. It is also worth noticing that for strong disorder one enters a regime where $n_0/n$ is significantly larger than $\rho_s/\rho$. More comprehensive results are shown in Fig.~\ref{fig4} where $n_c$ is estimated from the superfluid fraction, which is less sensitive to finite-size effects. The results clearly show an increase of the critical density as a function of $V_0$, from the non-interacting degenerate density $n_c^0$ up to values $\sim 20$ times larger. It is also worth noticing that for strong disorder, an increase of the scattering length $a$ is accompanied by a decrease of $n_c$ resulting in a constant value of the critical chemical potential (see Fig.~\ref{fig3}).

In conclusion, we have investigated the superfluid critical behavior of an interacting Bose gas in a correlated random medium. In the regime of strong disorder and low temperatures we identify a phase, where the gas is both normal and highly degenerate, which should be related to the Bose glass phase predicted at $T=0$. An important question that will be addressed in future studies concerns the equation of state and the thermodynamic properties of this exotic normal phase.

We acknowledge useful discussions with B. Svistunov, M. Modugno and L.P. Pitaevskii. This work, as part of the European Science Foundation EUROCORES Program ``EuroQUAM-FerMix'', was supported by funds from the CNR and the EC Sixth Framework Programme. NP acknowledges support from NSF grant PHY-0653183.
Calculations have been performed on the HPC facility {\it Wiglaf} at the Physics Department of the University of Trento and on the BEN cluster at ECT$^{\ast}$ in Trento.

\end{document}